\documentclass[aps,prl,twocolumn,floatfix,english,showpacs,10pt,superscriptaddress]{revtex4-1} \pdfoutput=1
\usepackage{amsmath,bm}
\usepackage{mathtools}
\usepackage{commath}
\usepackage{amssymb}
\usepackage{graphicx}
\usepackage{babel}
\usepackage{cases}
\usepackage[colorlinks=true, citecolor=blue, anchorcolor=green]{hyperref}
\hypersetup{linkcolor=red}
\usepackage{natbib}
\usepackage{floatrow}
\usepackage{dblfloatfix}
\usepackage{xcolor}
\usepackage{braket}

\makeatletter

\newcommand{\Rmnum}[1]{\expandafter\@slowromancap\romannumeral #1@}

\newcommand{\ex}{\text{ex}}

\newcommand{\eff}{\text{eff}}

\makeatother

\begin{document}
\title{Free-fermion multiply-excited eigenstates and their experimental signatures in 1D arrays of two-level atoms}
\author{Yu-Xiang Zhang}
\email{iyxz@nbi.ku.dk}
\affiliation{The Niels Bohr Institute, University of Copenhagen, Blegdamsvej 17, 2100 Copenhagen {\O}, Denmark}
\author{Klaus M\o lmer}
\affiliation{Aarhus Institute of Advanced Studies,  Aarhus University and Center for Complex Quantum Systems (CCQ), Department of Physics and Astronomy, Aarhus University\\8000 Aarhus C, Denmark}
\date{\today}

\begin{abstract}
One-dimensional (1D) subwavelength atom arrays display multiply-excited subradiant eigenstates which are reminiscent of free fermions. 
So far, these states have been associated with subradiant states with decay rates $\propto N^{-3}$, with $N$ the number of atoms, which fundamentally prevents detection of their fermionic features
by optical means. 
In this Letter, we show that free-fermion states generally appear
whenever the band of singly-excited states has a
quadratic dispersion relation at the band edge and may hence also be obtained with radiant and even superradiant states.  
1D arrays have free-fermion multiply-excited eigenstates that are typically either subradiant or (super)radiant, and we show that a simple transformation acts between the two families. Based on this correspondence, we propose different means for their preparation and analyze their experimental signature in optical detection.   
\end{abstract}
\maketitle

Subwavelength atom arrays support extremely subradiant states 
and significant optical nonlinearity~\cite{Chang:2018aa}, which bring promises
for photon 
storage ~\cite{Manzoni:2018aa,Asenjo-Garcia2017}, lossless mirrors~\cite{Rui:2020aa} and quantum
metasurfaces~\cite{Bekenstein:2020aa}, etc. 
Collective effects in subwavelength atom arrays can be described by the resonant 
dipole-dipole interaction
(RDDI) mediated by the quantized radiation field. The RDDI Hamiltonians
are long-range flip-flop spin models, and their
multiply-excited eigenstates display strong spatial correlations.
Such states have been obtained through studies
of 1D atom arrays coupled to waveguides~\cite{Asenjo-Garcia2017,Albrecht2018,Zhang:2019aa,Zhang:2020aa,Poddubny:2020aa,Rupasov:1984vo,Mahmoodian:2020aa,Iversen:2021vv,Ke:2019aa,Ke:2020vl,Zhong:2020uc,Iorsh:2020vd,Poshakinskiy:2021wu,Poshakinskiy:2021uq,Douglas:2016aa,Ruostekoski:2016aa,Ruostekoski:2017aa,Lang:2020vo}
and to the free space vacuum~\cite{Asenjo-Garcia2017,Manzoni:2017wd,Masson:2020ws,Javanainen:1999we,Williamson:2020vo,Williamson:2020vg,Cidrim:2020ub,Masson:2020ww,Mirza:2016tl}. 
For the waveguide case
where the RDDI Hamiltonian has a simple form~\cite{Chang2012}, a variety of 
spatial correlations of the two-excitation eigenstates have been obtained, see a summary of results in Ref.~\cite{Sheremet:2021uu}. Slater determinants formed by singly-excited eigenstates, so-called
free-fermion states~\cite{Asenjo-Garcia2017}, are regarded as a generic family of 
subradiant states in 1D atom arrays which is expected to appear also for
RDDI mediated by other fields, e.g., the free space vacuum field~\cite{Rui:2020aa}, 
and fields supported by hyperbolic metamaterials~\cite{Newman:2018aa}
or photonic crystal waveguides~\cite{Hood:2016vd}, etc.
Within the ideal 1D waveguide model
~\cite{Zhang:2019aa}, we have shown a mapping of the RDDI Hamiltonian to the Lieb-Liniger model~\cite{Lieb1963}
and interpreted the free-fermion subradiant states as a Tonks-Girardeau gas of hard-core bosons~\cite{Tonks1936,Girardeau1960}. 

However, two questions about the free-fermion states have remained unresolved.
First, a conclusive proof has not been given for their general existence.
Second, there has been a lack of methods for their detection, because the optical emission from the free-fermion subradiant states 
is suppressed by a factor of $N^{-3}$~\cite{Asenjo-Garcia2017}, with $N$ the number of atoms.

In this Letter, the first question is addressed by a
system-independent approach. We find that if the band of singly-excited states induced by 
RDDI has a quadratic extremum point $k_\ex$, i.e., its dispersion relation can be expanded as $\omega_{\eff}(k)\approx \omega_\eff(k_{\ex})+ a_2(k-k_\ex)^2$ for $k\approx k_\ex$ with $a_2$
an expansion coefficient, there is
a family of free-fermion multiply-excited eigenstates 
defined in the vicinity of $k_\ex$. This result provides a sufficient condition for
the generic existence of the free-fermion eigenstates and it
dismisses the notion that they must be extremely subradiant: There
exist free-fermion states with finite decay rates  and this paves new ways for their experimental detection. We thus propose two schemes to prepare and detect radiant (and even superradiant) free-fermion multiply-excited eigenstates of 1D atom arrays.

\paragraph*{Preliminaries.} We consider atoms with two levels, the 
ground state $\ket{g}$ and an excited state $\ket{e}$, between which the energy gap is $\omega_0$ ($\hbar=1$). 
In a regular 1D array, the atoms are equally spaced with coordinates $z_m=md$. The light field can be specified by
its dyadic Green's tensor $\mathbf{G}$. Assuming the Born-Markov approximation and translation symmetry of the light field in the direction along the array, 
the effective RDDI Hamiltonian is expressed as~\cite{Gruner1996,Dung1998,Dung2002}
\begin{equation}\label{heff}
H_\eff=-\mu_0\omega_0^2\sum_{m,n=1}^{N}
\mathbf{d}_m^{*}\cdot\mathbf{G}(z_m-z_n,\omega_0)\cdot\mathbf{d}_n\sigma_m^\dagger\sigma_n,
\end{equation}
where $\mu_0$ denotes the vacuum permeability, $\mathbf{d}$ is the transition
dipole moment, and  $\sigma^{\dagger}=\ket{e}\bra{g}$. The Hamiltonian~\eqref{heff} can be
rewritten in Fourier space as
\begin{equation}\label{heff-wk}
H_\eff={Nd}\int_{-\pi/d}^{\pi/d}\frac{dk}{2\pi}\;\omega_{\eff}(k) \;\sigma_k^{\dagger}\sigma_k,
\end{equation}
where the spin-wave operator reads
\begin{equation}
\sigma_k^{\dagger}=\frac{1}{\sqrt{N}}\sum_{m=1}^{N}e^{ikz_m}\sigma_m^{\dagger},
\end{equation}
and $\omega_{\eff}(k)$ is the complex dispersion relation of the band of singly-excited eigenstates.
For an infinite array, the state $\ket{k}\equiv\sigma_k^{\dagger}\ket{G}$, with $\ket{G}$ the atomic ground state,
has the energy $\Re\omega_{\eff}(k)$ and decay rate $\gamma(k)=-2\Im\omega_\eff(k)$,
where $\Re$ and $\Im$ denote the real and imaginary parts, respectively.

According to Eq.~\eqref{heff-wk}  $H_\eff$ 
is fully specified by the 
single excitation dispersion relation $\omega_\eff(k)$. Thus,
any generic feature of its eigenstates must reflect a common mathematical
property of $\omega_\eff(k)$. We note that for finite N, Eq.~\eqref{heff-wk} does not diagonalize the
Hamiltonian~\eqref{heff} because of non-trivial  spin commutator relations $[\sigma_k, \sigma_{k'}^{\dagger}]\neq \delta(k-k')$.
This leads to a rich variety of multiply-excited states
and the free-fermion state is only one of them, see, e.g., recent work on
atom arrays coupled to a 1D waveguide~\cite{Sheremet:2021uu}. The free-fermion states, 
however, are special as they exist for any Hamiltonian~\eqref{heff} as long as its
$\omega_\eff(k)$ has a quadratic bandedge $k_{\ex}$ where 
$\omega_{\eff}(k)\approx \omega_\eff(k_{\ex})+ a_2(k-k_\ex)^2$.

\paragraph*{Free-fermion states.} 
We shall show that $H_\eff$ can be approximated
(near the band edge) by a simpler Hamiltonian  
\begin{equation}\label{h1}
\mathbf{H}_1=c_1\hat{N}_e-\frac{a_2}{d^2}\sum_{j=1}^{N-1} 
\bigg(e^{-ik_\ex d}\, \sigma_j^{\dagger}\sigma_{j+1}+
 e^{ik_\ex d}\, \sigma_{j+1}^{\dagger}\sigma_{j} \bigg),
\end{equation}
where $c_1=\omega_{\eff}(k_\ex)+2a_2/d^2$ and $\hat{N}_e = \sum_{j=1}^{N}\sigma_j^{\dagger}\sigma_j$.
The dispersion relation of $\mathbf{H}_1$ is $\omega_1(k)=c_1-2a_2/d^2\cos[(k-k_\ex)d]$,
which equals $\omega_\eff(k)$ of $H_\eff$ near $k_\ex$.
$\mathbf{H}_1$ can be exactly diagonalized by the Jordan-Wigner transformation~\cite{Jordan:1928aa}
$\sigma_j^{\dagger}=e^{i\pi\sum_{m=1}^{j-1}f_m^{\dagger} f_m}f_j^{\dagger}$
and its Hermitian conjugate, where $f_m$ and $f_m^\dagger$
are fermionic operators satisfying the anti-commutation relations $\{ f_i, f_j^{\dagger}\}=\delta_{i,j}$ and $\{ f_i, f_j\}=0$. 
The transformation leads to
\begin{equation}
\mathbf{H}_1=\sum_{\xi=1}^{N} \omega_1(k_\ex+q_\xi)\; f_{\xi}^{\dagger} f_{\xi},
\end{equation}
where $f_{\xi}=\sum_{j=1}^{N}\braket{\psi_{\xi}|\sigma_j^{\dagger}|G}\,f_j$ is the 
annihilation operator for the single-excitation orthonormal mode
\begin{subequations}\label{basis}
\begin{equation}\label{1e}
\ket{\psi_\xi}=\frac{1}{\sqrt{2}} (\sigma^{\dagger}_{k_\ex+ q_{\xi}}-\sigma^{\dagger}_{k_\ex- q_\xi})\ket{G},
\end{equation}
with 
\begin{equation}\label{qxi}
q_\xi=\xi\frac{\pi/d}{N+1},\quad 1\leq\xi\leq N.
\end{equation}
\end{subequations}
An eigenstate of $\mathbf{H}_{1}$ with $n_e$ excitations has the form of
\begin{equation}\label{fermi-state}
\ket{F_{\vec{\xi}}}=f^{\dagger}_{\xi_1}f^{\dagger}_{\xi_2}\cdots f^{\dagger}_{\xi_{n_e}}\ket{G}.
\end{equation}
where $\vec{\xi}$ denotes the string $\xi_1\leq\xi_2\leq \cdots\leq\xi_{n_e}$.
The state lives in the vicinity of $k_{\ex}$ if $\xi_{n_e}\ll N$.

The idea of studying $H_\eff$ using a simpler Hamiltonian was 
recently used to prove a  power-law scaling of the decay rates of the singly-excited subradiant states~\cite{Zhang:2020ab}. As in~\cite{Zhang:2020ab}, to prove that
$\mathbf{H}_1$ approximates $H_\eff$, the idea is to write $H_\eff = \mathbf{H}_1 + \Delta H$ and show that 
$\Delta H$ can be treated as a perturbation to $\mathbf{H}_1$.
To proceed, we write $\Delta H$ in the form of Eq.~\eqref{heff-wk} with a
dispersion relation $\delta\omega(k)=\omega_{\eff}(k)-\omega_1(k)$
and evaluate the perturbative expression
\begin{equation}\label{F-wk}
\braket{F_{\vec{\xi}} | \Delta H | F_{\vec{\xi}} }
=Nd\int_{-\pi/d}^{\pi/d}\frac{\mathrm{d}k}{2\pi} \delta\omega(k) \braket{F_{\vec{\xi}} 
|\sigma_k^\dagger \sigma_k | F_{\vec{\xi}} }.
\end{equation}
By definition, $\delta\omega(k)$ scales as $N^{-3}$ for $\abs{k-k_\ex}\sim N^{-1}$, 
but generally as $O(1)$ outside the neighborhood of $k_\ex$.
(We assume that $\omega_{\eff}(k_\ex)$ is not degenerate with $\omega_{\eff}(k)$ at other wave numbers, as hybridization of these states may require further treatment.) 
We derive in the Supplemental Material~\cite{sp} 
that the occupation $\braket{F_{\vec{\xi}} | \sigma_k^\dagger \sigma_k | F_{\vec{\xi}} }$ scales as $O(1)$ for $\abs{k-k_\ex}\sim N^{-1}$ and as $N^{-4}$ elsewhere. 
Eq.~\eqref{F-wk} thus yields the scaling
$\braket{F_{\vec{\xi}} | \Delta H | F_{\vec{\xi}} } \propto N^{-3}$, which is a factor $N$ smaller than the separation of the eigenvalues of $\mathbf{H}_1$. 
The same scaling also holds for off-diagonal terms
$\braket{F_{\vec{\xi}} | \Delta H | F_{\vec{\xi'}}}$ where $\vec{\xi}\neq\vec{\xi'}$.
Therefore, $\Delta H$ can be consistently viewed as a perturbation to $\mathbf{H}_1$,
and $\ket{F_{\vec{\xi}}}$ are the leading order eigenstates of $H_{\eff}$.

Our result solidifies the following physical argument: A quadratic $\omega_\eff(k)$ 
corresponds to a kinetic energy that can be represented by  $\propto(\partial_x)^2$. Discrete versions of this operator reduce to nearest-neighbor tunneling. Therefore, although displaying long-range hopping terms, $H_{\eff}$ can be approximated by $\mathbf{H}_1$ and give rise to Jordan-Wigner fermions.

\paragraph*{Experimental signatures.} No assumption about subradiance is applied above. 
Radiant and even superradiant
free-fermion states are obtained if $\abs{k_{ex}}$ is smaller
than $k_0=\omega_0/c$ ($c$ is the speed of light). 
Within the Markov approximation, the
electric field (positive frequency part) of the emission from the atoms reads
\begin{equation}\label{E}
\hat{E}^{(+)}(\mathbf{r})=\mu_0\omega_0^2 \sum_{j=1}^{N}
\mathbf{G}(\mathbf{r}-\mathbf{r}_j, \omega_0)\cdot \mathbf{d}_j \sigma_j(t).
\end{equation}
In the far field, $\mathbf{G}(\mathbf{r},\omega_0)\propto r^{-1} e^{i k_0 r}\mathbf{f}(\theta,\phi)$, where $\mathbf{f}(\theta, \phi)$ is the
radiation pattern at polar angle $\theta$ and azimuthal angle $\phi$~\cite{Jackson:1999uq}. 
Thus, $\hat{E}^{+}(\theta)\propto \sqrt{N}\sigma_{k_0{\cos\theta}}$ and
quantities in the form of $\langle \sigma^{\dagger}_k\sigma_k\rangle$ 
and $\langle \sigma^{\dagger}_{k}\sigma^{\dagger}_{q}\sigma_q\sigma_k\rangle$, etc.,
can be efficiently measured if $k,q\in[-k_0, k_0]$. 
We thus propose two experimental schemes for the study of signatures of the free-fermion states. 

\paragraph*{Detection scheme-1.} First, we note that 1D atom arrays
usually have two bandedges, which are $k_{\ex,0}=0$
and $k_{\ex,\pi}=\pi/d$ if the system satisfies parity symmetry $\omega_{\eff}(k)=\omega_\eff(-k)$.
The implied free-fermion states are denoted by 
$\ket{F_{\vec{\xi}}^{0}}$ and  $\ket{F_{\vec{\xi}}^{\pi}}$, respectively. 
The states $\ket{F_{\vec{\xi}}^{\pi}}$ have been previously recognized as subradiant states when
$d<\pi/k_0$, while states $\ket{F_{\vec{\xi}}^{0}}$ are radiant and even superradiant. A one-to-one
correspondence between members of these two families 
is established by a single unitary
$U_{\pi}=\otimes_{m=1}^{N}(\ket{e_m}\bra{e_m}+(-1)^m \ket{g_m}\bra{g_m})$ so that
$U_\pi\ket{F^{0}_{\vec{\xi}}} =\ket{F^{\pi}_{\vec{\xi}}}$ and 
$U_\pi \ket{F^{\pi}_{\vec{\xi}}} =\ket{F^{0}_{\vec{\xi}}}$ for any $\vec{\xi}$.
$U_\pi$ factorizes and can be realized, e.g., by geometric phase control~\cite{He:2020vv}.

Subradiant states are difficult to excite directly by external lasers. We  can instead initialize the array in the symmetrically excited state, e.g.,  $\ket{\Psi_{0}}=\ket{B_{0,0,0}}\propto 
(\sigma_{k=0}^{\dagger})^3\ket{G}$ (see preparation methods discussed in Ref.~\cite{Albrecht2018}), and subsequently apply $U_\pi$ to transfer the excitations to $k_{\ex,\pi}$. The resulting state does not exclusively populate a free-fermion state, but it may be realized through a subsequent ``evaporative cooling'' process: Given no emission is observed,
the atomic state follows the ``no-jump'' trajectory  $\ket{\Psi_t}\propto e^{-iH_{\eff}t}U_{\pi}\ket{\Psi_0}$ and gradually converges to the most long-lived component in the eigenstate expansion of $U_{\pi}\ket{B_{0,0,0}}$. This state, in turn, is dominated by the desired free-fermion state, $\ket{F^{\pi}_{1,2,3}}$ 
(note that atom arrays coupled to a 1D waveguide may result in bound states with even longer lifetime~\cite{Zhang:2020aa,Poddubny:2020aa}).

We can arrest the ``cooling'' at any time and apply $U_\pi$ to convert the system to a radiant
state around $k_{\ex,0}$. 
Importantly, the fact that $U_{\pi}$ maps uniformly between $\ket{F_{\vec{\xi}}^{0}}$ and
$\ket{F_{\vec{\xi}}^{\pi}}$ for any $\vec{\xi}$ implies little 
deformation of the state during preparation. Then 
the time evolution is governed by the radiative master equation
\begin{equation}\label{master}
i\frac{d}{dt}\rho= {H}_{\eff}\rho-\rho\,{H}_{\eff}^{\dagger}
+i\sum_{\xi=1}^{N} \gamma_{\xi}\,\sigma_{\phi_\xi}\, \rho\,\sigma^{\dagger}_{\phi_\xi}
\end{equation}
where $\sigma_{\phi_\xi}=\sum_{j=1}^{N}\braket{\phi_{\xi}|\sigma_j^\dagger | G}\sigma_j$,
$\gamma_{\xi}$ and $\ket{\phi_\xi}$ are eigenvalues and eigenstates
of $2 H_{\eff}^{\operatorname{Im}}$, the dissipative part of $H_{\eff}$ defined through
$H_{\eff}\equiv H_{\eff}^{\operatorname{Re}}-iH_{\eff}^{\operatorname{Im}}$~\cite{Moelmer1993}, and the wave number subscript $\xi$ is counted with respect to $k_{\ex,0}$ in the manner of Eq.~\eqref{1e}.
Since $k_{\ex,0}$ is also a quadratic bandedge of $\Im\omega_{\eff}(k)$, we have $\ket{\phi_{\xi}}\approx \ket{\psi_{\xi}}$ for $\xi\ll N$.

\begin{figure}[bt]
\centering
\includegraphics[width=0.95\textwidth]{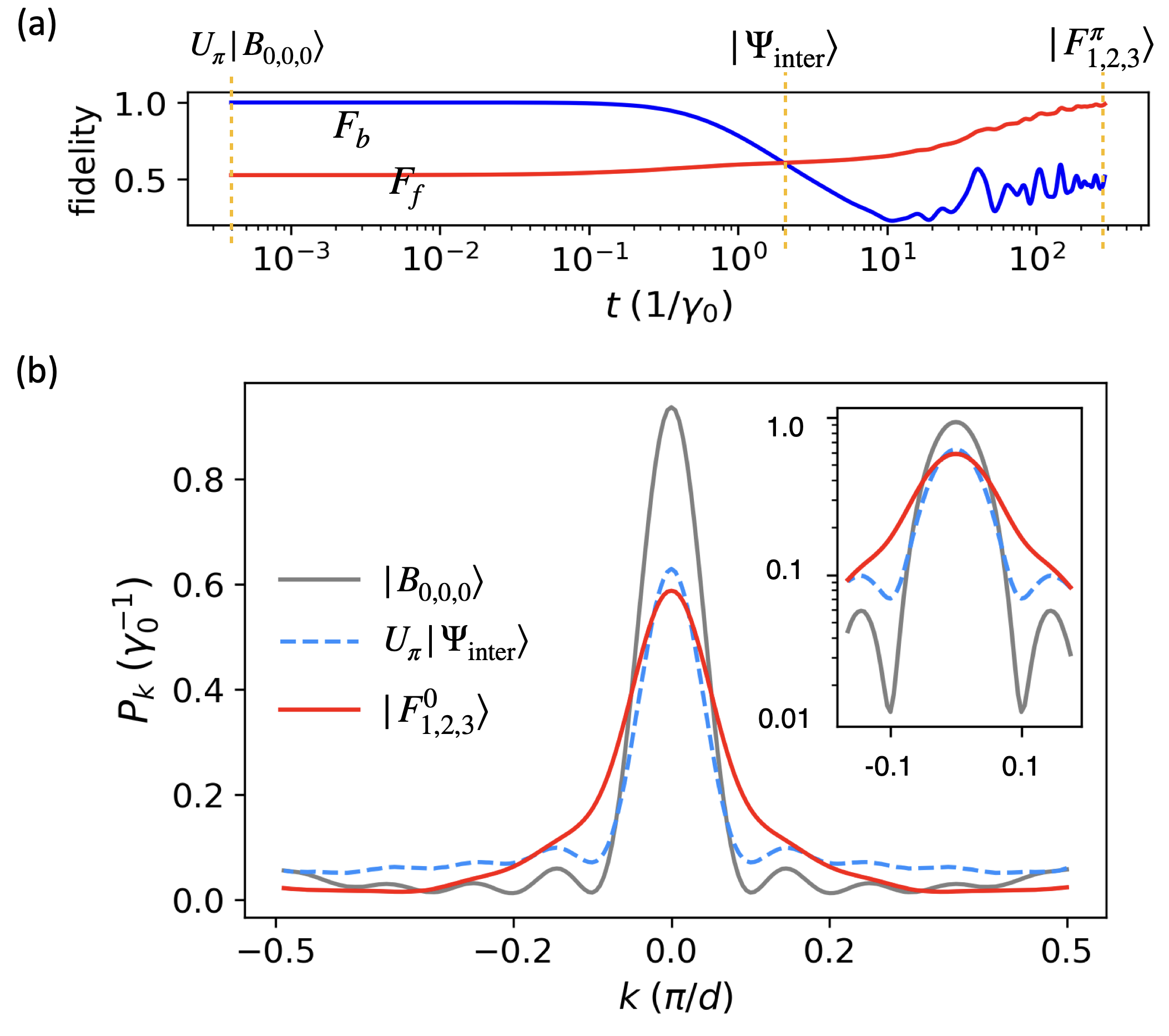}
\caption{Detection scheme-1. (a) Dissipative state evolution between the triply excited states $U_\pi \ket{B_{0,0,0}}$ and $\ket{F^{\pi}_{1,2,3}}$ of 20 atoms. Fidelities between $\ket{\Psi_t}$ and $U_{\pi}\ket{B_{0,0,0}}$ ($F_b$, blue) and $\ket{F^{\pi}_{1,2,3}}$ ($F_f$, red) as a function of time, in unit of $1/\gamma_0$ where $\gamma_0$ is the single atom spontaneous emission rate in 3D free space. The intermediate state $\ket{\Psi_{\text{inter}}}$  has
$F_b=F_f$. 
(b) Renormalized photon axial momentum distribution $P_k$ for $\ket{B_{0,0,0}}$, $U_\pi\ket{\psi_{\text{inter}}}$ and $\ket{F^{0}_{1,2,3}}$, respectively. The log scale inset shows
the central parts of $P_k$. 
}\label{fig1}
\end{figure}

We consider a 1D atom array
in vacuum in 3D space with $d=\lambda_0/4$ ($\lambda_0=2\pi/k_0$) and $N=20$, where
the atomic transition dipoles $\mathbf{d}$ are aligned parallel to the array~\cite{sp}. 
Such systems can be realized with 
sub-wavelength optical lattices~\cite{Ritt:2006wk,Rusconi:2021ta,Nascimbene:2015un,Anderson:2020tq,Olmos:2013aa}.
We refer the reader to Ref.~\cite{Rusconi:2021ta} for 
a thorough discussion of the influence of atomic motion, which is ignored here.
In Fig.~\ref{fig1}(a), we simulate the ``evaporative cooling'' process and plot
the fidelities $F_b=|\braket{\Psi_t|U_{\pi}|B_{0,0,0}}|^2$ (blue line) 
and $F_f=|\braket{{\Psi_t}| F^{\pi}_{1,2,3} }|^2$ (red line), respectively.
Along the no-jump trajectory, the atomic state coincides with the symmetrically excited and then phase flipped free-boson state $U_{\pi}\ket{B_{0,0,0}}$, an intermediate state $\ket{\Psi_{\text{inter}}}$ 
with equal overlap with $U_{\pi}\ket{B_{0,0,0}}$ and $\ket{F^{\pi}_{1,2,3}}$, and, finally, the desired free-fermion state $\ket{F^{\pi}_{1,2,3}}$. These three states are acted upon by $U_{\pi}$ and 
then used as the
initial state for the simulation of radiative emission governed by~\eqref{master}. 
In Fig.~\ref{fig1}(b), we plot the renormalized axial photon momentum distribution
$P_k=\int_{t}^{\infty} d\tau \langle \sigma_k^{\dagger}(\tau)\sigma_k(\tau)\rangle$, integrated over time. 
The distribution is defined according to Eq.~\eqref{E}, and evaluated by
averaging over 1000 quantum trajectories~\cite{Moelmer1993}.

States $\ket{F^{0}_{1,2,3}}$ and $\ket{B_{0,0,0}}$ are not orthogonal.
Fully populating one implies a 0.53 population of the other. However, as shown in Fig.~\ref{fig1}(b), 
the emission profile $P_k$ of
$\ket{F^{0}_{1,2,3}}$ is different from that of $\ket{B_{0,0,0} }$ by lower peak value and wider shoulders.
The insert in Fig.~\ref{fig1}(b) emphasizes the destructive interference 
in $P_{k}$ for  $\ket{B_{0,0,0}}$  at axial photon momentum $k\approx\pm 0.1\pi/d$.
The intermediate state $U_{\pi}\ket{\Psi_{\text{inter}}}$ overlaps equally with
$\ket{B_{0,0,0}}$ and $\ket{F^{0}_{1,2,3}}$, and shares emission features of both states.

\paragraph*{Detection scheme-2.} An alternative scheme may employ continuous laser excitation with a constant spatial phase, driving excitations with $k \simeq k_{\ex,0}$ of the atoms. Such driving is also studied in Refs.~\cite{Bettles:2016aa,Cidrim:2020ub,Holzinger:2021wa}.
The collective driving is modeled by
\begin{equation}\label{drive}
H_{L}=\Omega(e^{-i\delta_L t}\sigma^{\dagger}_{k=0}+e^{i\delta_L t}\sigma_{k=0})
\end{equation}
and we assume the detuning $\delta_L=\omega_0+\Re\omega_{\eff}(k_{\ex,0})$ so that $\ket{F_{1,2}}$ is the doubly-excited state
closest to resonance. The amplitude $\Omega$ is assumed to be weak
so that excited state components with $n_{e}\geq 3$ are neglected. The emitted radiation signal may be dominated by the most populated singly-excited components 
of the steady state, but we can extract the properties of the doubly-excited components by observation of photon coincidences, described by the 2nd-order equal-time correlation function 
$G(k_1,k_2)\equiv N^2\braket{\sigma_{k_1}^{\dagger}\sigma_{k_2}^{\dagger}\sigma_{k_2}\sigma_{k_1}}$.

If the RDDI is negligible, the two-excitation component of the steady state will be
$\ket{B_{0,0}}\propto (\sigma_{k=0}^{\dagger})^2\ket{G}$, and  only for sufficiently strong interactions
will the steady state, and hence the optical emission show features of $\ket{F_{1,2}}$. To focus on the essential physical mechanisms rather than
system-dependent details, we propose a minimal model
\begin{equation}\label{master-2}
i\frac{d}{dt}\rho=\mathbb{H}_1\rho-\rho\mathbb{H}_1^{\dagger}-[\rho, H_L]+i\beta\gamma_{\ex}
\sum_{\xi=1}^{N}\sigma_{\psi_\xi}\rho\sigma_{\psi_\xi}^{\dagger},
\end{equation}
where $\mathbb{H}_1$ is derived from $\mathbf{H}_1$,
\begin{equation}\label{hb1}
\mathbb{H}_1=\beta\frac{\gamma_\ex}{2i}\hat{N}_e-\frac{\Re(a_2)}{d^2}\sum_{j=1}^{N-1}
(\sigma_j\sigma_{j+1}^{\dagger}+\sigma_{j+1}\sigma_{j}^{\dagger}).
\end{equation}
In this model $\Delta\omega=\Re{a_2}/(Nd)^2$ characterizes the energy gaps between the
free-fermion states and $\beta\gamma_\ex$ characterizes their linewidths, where $\beta$ is introduced to
explicitly control the value of  the dimension-less ratio
$r_\beta=\Delta\omega/(\beta\gamma_\ex)$. When $r_\beta$ is large, eigenstates other than $\ket{F_{1,2}}$ are far from
resonance and the doubly-excited states predominantly occupy $\ket{F_{1,2}}$. Moreover, the quantum jump operators in Eq.~\eqref{master-2} appear with the same magnitude. Thus they
are equivalent to individual atomic decays. Eq.~\eqref{master-2} hence describes atoms decaying independently with decay rate $\beta\gamma_\ex$ while being coherently coupled to the nearest neighbors with ``renormalized'' tunneling strength $\Re a_2/d^2$.

\begin{figure}[tb]
\centering
\includegraphics[width=1\textwidth]{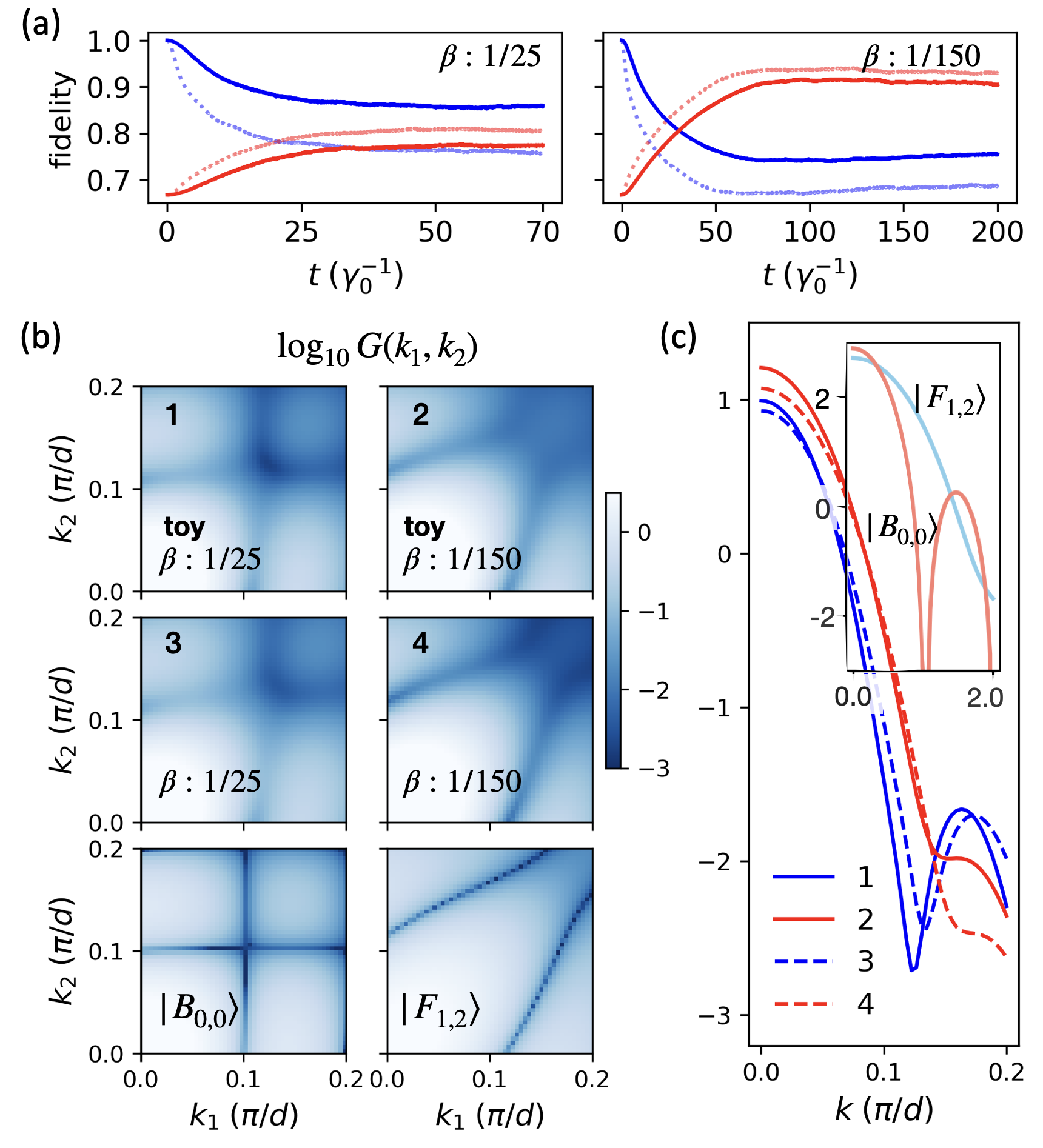}
\caption{Scheme-2. (a) Fidelities between the two-excitation components
of the steady states (normalized) and the bosonic state $\ket{B_{0,0}}$ (blue) and $\ket{F_{1,2}}$ (red). Solid lines
are for Eq.~\eqref{master-2} and dotted lines are for Eq.~\eqref{master}.
(b) Two-photon correlation function $\log_{10}G(k_1, k_2)$ evaluated in the steady states, with the 1st row for the steady states of Eq.~\eqref{master-2},
the 2nd row for steady states of Eq.~\eqref{master}, and the third row for $\ket{B_{0,0}}$ and $\ket{F_{1,2}}$. 
The color map on the right applies to the 1st and 2nd rows. Color map of the 3rd row is not shown.
(c) $\log_{10}G(k, k)$, i.e., values along the diagonal in the plots in (b). 
}\label{fig2}
\end{figure}

A simulation of Eq.~\eqref{master-2} is compared with a simulation
based on Eq.~\eqref{master}, including $H_L$, for the same system and parameters of $H_{\eff}$ as used in Scheme-1 but allowing variation of the dissipative part through the  factor $\beta$, i.e.,
$H_{\eff}\rightarrow H_{\eff}^{\operatorname{Re}}-i\beta H_{\eff}^{\operatorname{Im}}$. 
This choice of $H_{\eff}$ yields $\Re\omega_\eff(k_\ex)\approx-1.03\gamma_0$, 
$\gamma_{\ex}\approx 3\gamma_0$, and
$\Re a_2/d^2\approx 0.17\gamma_0$, that we apply in  Eq.~\eqref{master-2}. 
In Fig.~\ref{fig2} we show the results of the simulation for $N=20$ atoms
with paired parameters ($\beta=1/25$, $\Omega=0.01\gamma_0$) and
($\beta=1/150$, $\Omega=0.008\gamma_0$). 
In the Supplemental Material~\cite{sp}, we show results for  a larger value of $N=30$ which blurs some of the features of $\ket{F_{1,2}}$. The results are obtained by averaging over
1000 quantum trajectories~\cite{Moelmer1993}.

In Fig.~\ref{fig2}(a), we extract the two-excitation component of each quantum state trajectory,
renormalize it, and plot its fidelity with $\ket{F_{1,2}}$ and $\ket{B_{0,0}}\propto (\sigma_{k=0}^{\dagger})^2\ket{G}$, for both Eq.~\eqref{master-2} and Eq.~\eqref{master}.
We select two values of $\beta$, $1/25$ and $1/150$. For
 $\beta=1/25$, the steady state of  Eq.~\eqref{master}
(left panel, dotted lines, $r_\beta\approx 0.004$) is at
an intermediate stage with equal overlaps with
$\ket{F_{1,2}}$ and $\ket{B_{0,0}}$ while for Eq.~\eqref{master-2} (solid line) the dominant overlap is with
 $\ket{B_{0,0}}$. For $\beta=150$ (right panel, $r_\beta\approx 0.02$), both models yield
dominant overlap with the free-fermion state $\ket{F_{1,2}}$.

In Fig.~\ref{fig2}(b), we plot the two-photon coincidences, $\log_{10}G(k_1,k_2)$ with $0\leq k_{1}, k_2\leq 0.2\pi/d$
for the steady states of Eq.~\eqref{master-2} (the first row, labelled by ``toy'') 
and Eq.~\eqref{master} (the second row). In the plots, the patterns colored by dark blue represent suppression of two-photon coincidences. In each row, plots for
$\beta=1/25$ and $\beta=1/150$ are shown in the left and right column, respectively.
In the third row we show results evaluated as expectation values in the states $\ket{B_{0,0}}$ (left) and $\ket{F_{1,2}}$ (right).
 
The almost identical patterns obtained for the same $\beta$ in Fig.~\ref{fig2}(b) show that
Eq.~\eqref{master-2} approximates Eq.~\eqref{master} well. For
$\beta=1/25$ the patterns display an upright cross as a signature of the state $\ket{B_{0,0}}$.
While for $\beta=1/150$, we see two sloping lines characterizing $\ket{F_{1,2}}$.
To distinguish them more quantitatively,  in Fig.~\ref{fig2}(c) we plot the diagonal terms, i.e.,
$\log_{10}G(k, k)$, of the four subplots labeled ``1-4'' in Fig.~\ref{fig2}(b),  and plot those of 
$\ket{B_{0,0}}$ and $\ket{F_{1,2}}$ in the insert. 
The insert shows that the upright cross of $\ket{B_{0,0}}$ results in anti-bunching at $k\approx 0.1\pi/d$, 
while the avoided crossing of $\ket{F_{1,2}}$ 
leads to a more smooth curve. The anti-bunching  is clearly seen 
in the blue solid and dashed lines ($\beta=1/25$), but are smoothed in
the red lines ($\beta=1/150$).
The qualitative agreement between the dashed lines and solid lines in Fig.~\ref{fig2}(c) further validates the approximate treatment by  Eq.~\eqref{master-2}.
  
Candidate systems supporting large values of $r_\beta$ are atom arrays coupled to photonic crystals 
where the atomic transition frequency $\omega_0$ is in the vicinity of the  photonic band edge~\cite{Hood:2016vd,Gonzalez-Tudela:2015vz}; and 
atom arrays coupled to 1D waveguide modes, where at $k_{\ex}=0$
$H_{\eff}^{\operatorname{Re}}$ is enhanced while $H_{\eff}^{\operatorname{Im}}$ is 
reduced due to coupling via residual non-guided modes~\cite{Asenjo-Garcia2017}.
With a more sophisticated state preparation, 
one may distinguish free-fermion and free-boson ansatz states directly by the measurements. 
For example, for
$\ket{F^{0}_{\xi_1,\xi_2}}$ the correlation function $G(k, -k)=\langle \sigma_{-k}^\dagger\sigma_k^\dagger\sigma_k\sigma_{-k}\rangle$ will vanish for any $k$ as long as $\xi_1+\xi_2$ is an even integer, while for
$\ket{B_{\xi_1,\xi_2}}$ it vanishes when $\xi_1+\xi_2$ is odd.

\paragraph*{Discussions and Conclusions.}
To conclude, by approximating the RDDI Hamiltonian $H_{\eff}$ with the solvable model 
$\mathbf{H}_1$, the free-fermion states are found to be a generic consequence of 
the quadratic dispersion relation near the bandedge of the singly-excited states. 
We propose to observe the free-fermion states by their optical emission
in two basic schemes. Scheme-1 combines unitary and dissipative evolution to prepare a subradiant free-fermion state and exploits a transfer of the quantum system  between the sub- and super-radiant states
to observe the directional  distribution of radiation. 
Scheme-2 observes the 2nd-order correlation function of the
steady state emission by the atoms subject to constant laser driving. 

The free-fermion state is not a precise ansatz if the extremum point of $\omega_\eff(k)$ is not quadratic. 
Quartic extremum points exist in atom arrays in 3D free space~\cite{Zhang:2020ab}. In this case, 
$\mathbf{H}_1$~\eqref{h1} should be replaced by one with beyond nearest neighbor tunneling processes.
After the Jordan-Wigner transformation, this corresponds to a strongly-interacting fermionic model.

\begin{acknowledgements}
Y.-X. Zhang thanks Anders S{\o}rensen, Bj{\"o}rn Schrinski, and Johannes Bjerlin for valuable discussions. The authors acknowledge financial support from the Danish National Research Foundation through the Center  for Hybrid  Quantum  Networks (Grant Agreement DNRF 139) and the Center for Complex Quantum Systems (Grant Agreement DNRF 156). K.M also acknowledges European Union’s Horizon 2020 Research and Innovation Programme under the Marie Sklodowska- Curie program (754513).
\end{acknowledgements}

\bibliography{supsub}

\end{document}


\title{Supplemental Material to ``Free-fermion multiply-excited eigenstates and their experimental signatures in 1D arrays of two-level atoms''}
\author{Yu-Xiang Zhang}
\affiliation{The Niels Bohr Institute, University of Copenhagen, Blegdamsvej 17, 2100 Copenhagen {\O}, Denmark}
\author{Klaus Mølmer}
\affiliation{Aarhus Institute of Advanced Studies,  Aarhus University\\ Center for Complex Quantum Systems (CCQ), Department of Physics and Astronomy, Aarhus University\\8000 Aarhus C, Denmark}
\date{\today}

\begin{abstract}
In this Supplemental Material, we present the detailed proof of our main result that a  quadratic single excitation dispersion relation 
implies free-fermion multiply-excited states of 1D arrays of atomic dipoles in Sec.~\ref{proof}; 
we derive Eq.~(9) of the main text in Sec.~\ref{derive10};
we present elements of the theory of atomic arrays in 3D free-space and evidence of free-fermion superradiant states that are used in our detection scheme-1 in Sec.~\ref{scheme1}, and we show simulations of our detection scheme-2 with $N=30$ atoms
in Sec.~\ref{scheme2}.
\end{abstract}

\maketitle


\section{Quadratic dispersion implies free-fermion states}
\label{proof}

Here we present our main result in details. Without loss of generality, we assume $k_\ex=0$
throughout this section. Cases
of $k_{\ex}\neq 0$ can be obtained by a formal translation in the Brillouin zone.

\subsection{Write $\ket{F_{\vec{\xi}}}$ in the spin basis}
In the main text, the free-fermion state $\ket{F_{\vec{\xi}}}$ is expressed with the Jordan-Wigner fermions.
Here we write them in the spin basis for later convenience.

To start, we introduce a state of $n_e$ excitations, which are ordered in the 1D array and specified by a wavenumber
\begin{equation}\label{k}
\begin{aligned}
\ket{\vec{k_{\xi}}}& =\ket{k_{\xi_1}, k_{\xi_2},\cdots, k_{\xi_{n_e}}} \\
&=\frac{1}{(\sqrt{N})^{n_e}}\sum_{x\uparrow} e^{i\sum_{j=1}^{n_e} k_{\xi_j} x_j} \ket{x_1, x_2\cdots x_{n_e}}.
\end{aligned}
\end{equation}
Above we have used the basis in real space
\begin{equation}
\ket{x_1, x_2\cdots x_{n_e}}=(\bigotimes_{j=1}^{n_e}\sigma_{x_j}^{\dagger})\ket{G},
\end{equation}
with the short-hand notation
\begin{equation}
\sum_{x\uparrow}\cdots \equiv \sum_{z_1\leq x_1<x_2\cdots < x_n \leq z_N}\cdots,
\end{equation}
where $z_1=d$ and $z_N=Nd$ are coordinates of the first and last atom. Similarly, we will use the notation
\begin{equation}
\ket{\psi_1, \psi_2,\cdots \psi_{n_e}}=\sum_{x\uparrow}
\bigg[\prod_{j=1}^{n_e}\psi_j(x_j)\bigg]\ket{x_1,x_2\cdots, x_{n_e}}
\end{equation}
Then the free-fermion state $\ket{F_{\vec{\xi}}}$ is expressed as
\begin{equation}\label{fxi}
\begin{aligned}
\ket{F_{\vec{\xi}}} 
&=\sum_{\p}(-1)^{\p}\ket{\psi_{\p(\xi_1)}, \psi_{\p(\xi_2)}\cdots, \psi_{\p(\xi_{n_e})}}\\
&= \sum_{\p}(-1)^{\p} \sum_{x\uparrow}  \bigg[\prod_{j=1}^{n_e}\psi_{\p (\xi_j)}(x_j)\bigg]\ket{x_1,x_2,\cdots x_{n_e}} \\
&=\frac{1}{\sqrt{2^{n_e}}} \sum_{\p}(-1)^{\p} \bigg[\prod_{j=1}^{n_e}\epsilon_j\bigg]
\ket{\epsilon_1 k_{\p(\xi_1)}, \cdots \epsilon_{n_e} k_{\p(\xi_{n_e})}}
\end{aligned}
\end{equation}
where $(-1)^{\p}$ denotes the parity of the permutation of $\vec{\xi}$, 
$\psi_{\p(\xi_j)}(x_j)=\braket{G|\sigma_j|\psi_{\p(\xi_j)}}$ with $\ket{\psi_{\p(\xi_j)}}$ defined in Eq.~(6)
of the main text, which can be rephrased as
\begin{equation}
\ket{\psi_\xi}=\frac{1}{\sqrt{2}} (\ket{q_\xi} -\ket{-q_\xi}).
\end{equation}
In the 3rd equality of Eq.~\eqref{fxi}, $\epsilon_j=\pm 1$ for $j=1,2,\cdots, n_e$.

\subsubsection{Normalization of $\ket{F_{\vec{\xi}}}$}
We use the above notation to verify the normalization of $\ket{F_{\vec{\xi}}}$.
Consider the inner product
\begin{equation}
\braket{F_{\vec{\xi}}| F_{\vec{\zeta}}} 
=  \sum_{\p, \p'}(-1)^{\p\p'}\sum_{x\uparrow} \prod_{l=1}^{n}\psi^*_{\p (\xi_l)}(x_l)\psi_{\p' (\zeta_l)}(x_l).
\end{equation}
We can replace the summation of permutation $\p$ and $\p'$
by first choosing a $\p_0$ and sum over $\p$ and $\p'=\p\p_0$:
\begin{equation}
\begin{aligned}
&\braket{F_{\vec{\xi}}| F_{\vec{\zeta}}} \\
=&\sum_{\p_0,\p}(-1)^{\p_0}\sum_{x\uparrow} \prod_{l=1}^{n}\psi^*_{\p (\xi_l)}(x_l)\psi_{\p\p_0 (\zeta_l)}(x_l)\\
=&\sum_{\p_0,\p}(-1)^{\p_0}\sum_{x\uparrow}\prod_{l=1}^{n}\psi^*_{\xi_l}[\p^{-1}(x_l)]\psi_{\p_0 (\zeta_l)}[\p^{-1}(x_l)] \\
=& \sum_{\p_0}(-1)^{\p_0}\sum_{\p}\sum_{\p x\uparrow}\prod_{l=1}^{n}\psi^*_{\xi_l}(x_l)\psi_{\p_0 (\zeta_l)}(x_l),
\end{aligned}
\end{equation}
where in the 2nd equality, the permutation is moved from the subindex of $\psi$ to its coordinate argument;
and in the 3rd equality, the replacement $x_l\rightarrow \p (x_l)$ is used.

Next, we notice that now the following substitution is allowed:
\begin{equation}
\sum_{\p}\sum_{\p x\uparrow} \rightarrow \sum_{x_1=z_1}^{z_N}\sum_{x_2=z_1}^{z_N}\cdots \sum_{x_n=z_1}^{z_N}.
\end{equation}
Note that nonphysical contributions with $x_i=x_j$ cancel each other due to the anti-symmetry of the free-fermion states implied in our calculation.
Using the equality $\braket{\psi_{\xi_l} | \psi_{\xi_m}}=\delta_{l, m}$, we immediately obtain
\begin{equation}
\braket{F_{\vec{k}}| F_{\vec{q}}} 
=\sum_{\p_0}(-1)^{\p_0}\delta_{\vec{k},\p_0(\vec{q})}.
\end{equation}
This shows that the free-fermion states are orthonormal.

\subsection{An equality for $\sigma_k\ket{F_{\vec{\xi}}}$}
Here we derive an expansion of $\sigma_k\ket{F_{\vec{\xi}}}$, which will be used to estimate its magnitude.
The derivation is separated into a few steps.

\subsubsection{the evaluation of $\sigma_k\ket{\vec{q}}$}
Consider a state $\ket{\vec{q}}$ defined in the form of Eq.~\eqref{k}, which is a component of
$\ket{F_{\vec{\xi}}}$ given in Eq.~\eqref{fxi}. We have the following expansion
\begin{widetext}
\begin{equation}
\begin{aligned}
\sqrt{N}\sigma_k\ket{\tilde{q}} 
=&\sum_{j=1}^{N}\frac{1}{N^{n_e/2}}\sum_{x\uparrow}e^{i\sum_{l=1}^{n}q_lx_l-ikx_j}\sigma_j\ket{x_1, x_2\cdots, x_n} \\
=&\sum_{m=1}^{n_e}\frac{1}{N^{n_e/2}} \sum_{x\uparrow}e^{i\sum_{l=1}^{n}q_lx_l-ikx_m}
\ket{\cdots x_{m-1}, x_{m+1}\cdots x_n}.
\end{aligned}
\end{equation}
In the 2nd equality, the summation over $x_m$ can be evaluated using
the fact that $x_{m-1}<x_m<x_{m+1}$:
\begin{equation}
\sum_{x_m=x_{m-1}+d}^{x_{m+1}-d}
e^{i (q_m-k) x_m} =\frac{e^{i(q_m-k)(x_{m-1}+d)}-e^{i (q_m-k)x_{m+1}}}{1-e^{i (q_m-k) d}}.
\end{equation}
The above formula assumes $q_m\neq k$, while the case of $q_m=k$ can be recovered by taking
the limit of the further derived expressions as  $q\rightarrow k_m$. Finally, we have
\begin{equation}
\begin{aligned}
N\sigma_k\ket{\vec{q}} & =
\frac{e^{i({q}_1-k) z_1}}{1-e^{i({q}_1-k) d}}\ket{q_2,	q_3\cdots, q_n} - \frac{e^{i({q}_n-k) (z_N+d)}}{1-e^{i({q}_n-k) d}}\ket{q_1, q_2\cdots, q_{n-1}} \\
 &\quad +\sum_{m=1}^{n-1} f_k({q}_m, {q}_{m+1}) \ket{q_1,\cdots, q_{m-1}, q_m+q_{m+1}-k, q_{m+2},
\cdots, q_{n}} 
\end{aligned}
\end{equation}
where
\begin{equation}
f_k({q}_m, {q}_{m+1}) 
=-1+\frac{i}{2}\cot(\frac{q_{m+1}-k}{2}d)-\frac{i}{2}\cot(\frac{q_{m}-k}{2}d).
\end{equation}

\subsubsection{the calculation of $\sigma_k\ket{F_{\vec{\xi}}}$ }
We start from one particular sequence of arguments $\vec{\xi}$,
\begin{equation}\label{skFq_1}
\begin{aligned}
N\sigma_k\ket{\psi_{\xi_1}, \psi_{\xi_2}\cdots, \psi_{\xi_{n_e}}}
=&f_{k,L}(q_1)\ket{\psi_{\xi_2},\psi_{\xi_3}, \cdots, \psi_{\xi_{n_e}}}
-f_{k,R}(q_n) \ket{\psi_{\xi_1}, \psi_{\xi_2},\cdots, \psi_{\xi_{n_{e}-1}}}\\
+\sum_{m=1}^{n_e-1}\sum_{\epsilon_{1,2}=\pm} & \epsilon_1\epsilon_2
f_k(\epsilon_1{q}_m, \epsilon_2{q}_{m+1})\ket{\psi_{\xi_1},\cdots, \psi_{\xi_{m-1}},
\epsilon_1 q_m+ \epsilon_2 q_{m+1}-k, \psi_{\xi_{m+2}}, \cdots, \psi_{\xi_{n_e}}}
\end{aligned}
\end{equation}
\end{widetext}
where the chain-end coefficients are
\begin{subequations}
\begin{equation}
f_{k,L}(q_1)=\frac{e^{i(q_1-k) z_1}}{1-e^{i(q_1-k) d}}
-\frac{e^{i(-q_1-k) z_1}}{1-e^{i(-q_1-k) d}}
\end{equation}
\begin{equation}
f_{k,R}(q_n)=\frac{e^{i(q_n-k) (z_N+d)}}{1-e^{i(q_n-k) d}}-
\frac{e^{i(-q_n-k) (z_N+d)}}{1-e^{i(-q_n-k) d}}.
\end{equation}
\end{subequations}
Following Eq.~\eqref{fxi}, now
we consider the permutations and evaluate
\begin{equation}
N\sigma_k\sum_{\p} (-1)^{\p}\ket{\psi_{\p (\xi_1)},\cdots \psi_{\p (\xi_{n_e})}}.
\end{equation}
Substituting Eq.~\eqref{skFq_1} into the permutations, we see that
terms corresponding to the right hand side of the first lines of Eq.~\eqref{skFq_1} are
straightforward $(n_e-1)$-excitation states, but those in the second line are more involved with  $(n_e-2)$ arguments in the form of $\psi_{\xi}$ and one in the form of 
$\epsilon_1q_\alpha+\epsilon_2 q_\beta-k$. 
We start by collecting all terms that have the same group of arguments but
in different orders. We assume that $\alpha=m$ and $\beta=m+1$, and we fix a particular order of $(n_e-2)$ indices, 
\begin{equation}\label{string}
(\xi_1, \cdots, \xi_{m-1}, \xi_{m+2},\cdots \xi_{n_e}).
\end{equation}
Next, we insert $(\xi_{m}, \xi_{m+1})$, as a pair in one of the $(n_e-1)$ possible locations in the above string of $(n-2)$ indices.
This yields
$(n_e-1)$ different permutations of the string $\vec{\xi}$, which all have the same parity.
 
For every one of these $(n_e-1)$ permutations, one can apply
Eq.~\eqref{skFq_1} and obtain one term resembling the second line of Eq.~\eqref{skFq_1}, i.e.,
a state denoted with $\epsilon_1 q_m+\epsilon_2 q_{m+1}-k$ inserted in a string of $\psi$.
The sum of all these (n-1) terms is hence
\begin{widetext}
\begin{equation}\label{tempt-1}
\epsilon_1\epsilon_2 f_k(\epsilon_1q_m, \epsilon_2 q_{m+1})
\sigma^{\dagger}_{\epsilon_1 q_m+\epsilon_2 q_{m+1}-k} 
\ket{\psi_{\xi_1}\cdots \psi_{\xi_{n_e}}}_{m,m+1}
\end{equation}
where we have used a short-hand notation 
\begin{equation}\label{skFq_2}
\ket{\psi_{\xi_1}\cdots \psi_{\xi_{n_e}}}_{m,m+1}
=\ket{\psi_{\xi_1}\cdots, \psi_{\xi_{m-1}},\psi_{\xi_{m+2}},\cdots \psi_{\xi_{n_e}}},
\end{equation}
and used the fact that
\begin{equation}
\sigma_k^{\dagger}\ket{\vec{q}}= \ket{k,q_1, q_2\cdots}+\ket{q_1, k, q_2, q_3\cdots}
+\ket{q_1, q_2, k, q_3\cdots}+\cdots+\ket{q_1, q_2,\cdots, q_{n_e},k}.
\end{equation}
Next, these states can also be generated in the same way by inserting the ordered pair 
$(q_{m+1}, q_{m})$ into Eq.~\eqref{string}. The parity of these
permutations are just opposite to those obtained above.
By adding them together, we replace the factor $f_k(\cdots)$ in Eq.~\eqref{tempt-1} by
\begin{equation}\label{skFq_4}
\begin{aligned}
\lambda_k(\epsilon_1q_m, \epsilon_2 q_{m+1}) 
\equiv &f_k(\epsilon_1q_m, \epsilon_2 q_{m+1})-f_k(\epsilon_2q_{m+1}, \epsilon_1 q_{m}) \\
= & i\cot(\frac{k-\epsilon_1q_m}{2}d)-i\cot(\frac{k-\epsilon_2 q_{m+1}}{2}d).
\end{aligned}
\end{equation}
For any pair $(q_\alpha, q_{\beta})$, we introduce the notation
\begin{equation}
\sigma^{\dagger}_{q_\alpha,q_\beta;k}  =(-1)^{\alpha+\beta-1} \sum_{\epsilon_{\alpha},\epsilon_{\beta}=\pm 1}
\epsilon_\alpha\epsilon_\beta
\lambda_k(\epsilon_\alpha q_\alpha, \epsilon_\beta q_\beta ) 
 \sigma^{\dagger}_{\epsilon_\alpha q_\alpha+\epsilon_\beta q_\beta -k}
\end{equation}
and obtain the concise formula that
\begin{equation}\label{central-equation}
N\sigma_k\ket{F_{\vec{\xi}}}
= \sum_{\alpha=1}^{n_e}f_{k,LR}(q_{\alpha})\ket{F_{\vec{\xi}}}_\alpha
+\sum_{\alpha<\beta} \sigma^{\dagger}_{q_\alpha, q_\beta;k}\ket{F_{\vec{\xi}}}_{\alpha,\beta}.
\end{equation}
Here, $\ket{F_{\vec{\xi}}}_{\alpha}$ and
$\ket{F_{\vec{\xi}}}_{\alpha,\beta}$ are defined 
in the manner of Eq.~\eqref{skFq_2}, i.e., fermionic states defined
upon the ordered sequence $\vec{\xi}$ with $\xi_{\alpha}$ or both $\xi_{\alpha}$ and $\xi_{\beta}$, 
removed, respectively; and $f_{k,LR}$ reads
\begin{equation}\label{fklr}
\begin{aligned}
f_{k,LR}(q_{\alpha}) & \equiv (-1)^{\alpha+1}[f_{k,L}(q_\alpha)+(-1)^{n_e} f_{k,R}(q_\alpha)] \\
 & = (-1)^{\alpha+1}[1+(-1)^{n_e} e^{-i(q_\alpha -k)(N+1)d}]
\frac{i}{2}\bigg[\cot(\frac{k+q_{\alpha}}{2}d)-\cot(\frac{k-q_\alpha}{2}d) \bigg].
\end{aligned}
\end{equation}
In the second equality we have employed the
conditions that $e^{i2q_\alpha (N+1)d}=1$ and $z_1=d$.
\end{widetext}

\subsection{Eigenvalues of $\sigma_k^{\dagger}\sigma_k$ and $\sigma_k\sigma_k^{\dagger}$  }
Here we study the eigenvalues of $\sigma_k^{\dagger}\sigma_k$, which will be used in the next subsection.
First of all, we use a local unitary transformation
\begin{equation}
U_k=\bigotimes_{i=-\infty}^{+\infty}\bigg(e^{ik x_i}\ket{e_i}\bra{e_i}+\ket{g_i}\bra{g_i}\bigg).
\end{equation}
to transform $\sigma_k^{\dagger}\sigma_k$ into the standard collective spin operator
\begin{equation}
S^{\dagger}S= U_k^{\dagger}\sigma_k^{\dagger}\sigma_kU_k
\end{equation}
where $S^\dagger=\frac{1}{\sqrt{N}}\sum_{i=1}^{N} \sigma_i^\dagger$.
Then the problem is reduced to the spectrum of $S^{\dagger}S$, which
can be expressed by angular momentum operators:
\begin{equation}
S^\dagger S=\frac{4}{N}( L^2-L_z^2+L_z )
\end{equation}
where $L^2$ is the squared magnitude of the 
total angular momentum and $L_z=\frac{1}{2}\sum_i \sigma_i^{z}$ is the
z-component.

In the space of $n_e$-excitation states, the eigenvalue of $L_z$ is $l_z=-N/2+n_e$. 
The eigenvalue of $L^2$ is $l(l+1)$, where the values of $l$ 
should satisfy the relation that $\abs{l_z}\leq l \leq N/2$. 
Let us introduce $x=l-\abs{l_z}$. Then the eigenvalues of $S^\dagger S$ are expressed as
\begin{equation}
\frac{4}{N}x(N-2n_e+x+1),\quad 0\leq x\leq n_e.
\end{equation}
Given that $n_e\ll N$, the
magnitude of the eigenvalues of $S^{\dagger}S$ do not scale with $N$,
i.e., $\norm{S^\dagger S}\sim O(1)$. Similarly, the eigenvalues of $SS^\dagger=\frac{4}{N}(L^2-L_z^2-L_z)$ are
\begin{equation}
\frac{4}{N}(x+1)(N-2n_e+x),\quad 0\leq x\leq n_e,
\end{equation}
hence $\norm{SS^\dagger}\sim O(1)$ as well.

\subsection{A key inequality }
Here we estimate the magnitude of $\norm{\sigma_k\ket{F_{\vec{q}}}}$ by using the triangle inequality
on Eq.~\eqref{central-equation}
\begin{widetext}
\begin{equation}\label{skFq_3}
 \norm{\sigma_k\ket{F_{\vec{\xi}}}}
\leq  \frac{1}{N}\sum_{\alpha=1}^{n_e}\abs{f_{k,LR}(q_\alpha)} 
+\frac{1}{N}\sum_{\alpha<\beta}\norm{\sigma^{\dagger}_{q_\alpha, q_\beta;k}\ket{F_{\vec{\xi}}}_{\alpha,\beta}},
\end{equation}
where we have used the normalization $\norm{\ket{F_{\vec{\xi}}}_\alpha}=1$. 
There are totally $n_e(n_e+1)/2$ terms in the right hand side
of the inequality~\eqref{skFq_3}, but $n_e^2$ does not
scale with $N$, given that we restrict ourselves to 
$n_e\ll N$. 

From Eq.~\eqref{fklr}, it can be seen that the first term on the right hand side of the
inequality~\eqref{skFq_3} is bounded from above by
\begin{equation}
\begin{aligned}
\abs{f_{k,LR}(q_\alpha)} & \leq \abs{h_k(-q_\alpha, q_\alpha)} \\
&= \abs{\cot({\frac{k+q_\alpha}{2}d})
-\cot({\frac{k-q_\alpha}{2}d})},
\end{aligned}
\end{equation}
where 
$$h_k(q_1, q_2)=\cot[(k-k_\ex-q_1)d/2]-\cot[(k-k_\ex-q_2)d/2].$$

For the second term on the right hand side of the inequality~\eqref{skFq_3}, 
we notice that 
\begin{equation}\label{sigmaF}
\begin{aligned}
\norm{\sigma^{\dagger}_{q_\alpha, q_\beta;k}\ket{F_{\vec{\xi}}}_{\alpha,\beta}}
&\leq \sum_{\epsilon_{\alpha},\epsilon_{\beta}=\pm 1}\abs{\lambda_k(\epsilon_\alpha q_\alpha, \epsilon_\beta q_\beta)}
\norm{\sigma^{\dagger}_{\epsilon_\alpha q_\alpha+\epsilon_\beta q_\beta-k}\ket{F_{\vec{\xi}}}_{\alpha,\beta}} \\
&\leq \sum_{\epsilon_{\alpha},\epsilon_{\beta}=\pm 1}\abs{\lambda_k(\epsilon_\alpha q_\alpha, \epsilon_\beta q_\beta)}
\sqrt{\max_{n_e-2}\norm{\sigma_{\epsilon_\alpha q_\alpha+\epsilon_\beta q_\beta-k}
\sigma^{\dagger}_{\epsilon_\alpha q_\alpha+\epsilon_\beta q_\beta-k}}}
\end{aligned}
\end{equation}
where by $\max_{n_e-2}\norm{\cdots}$ we mean the largest eigenvalue in the subspace of $(n_e-2)$ excitations. 
We have
\begin{equation}
\max_{n_e-2}\norm{\sigma_{\epsilon_\alpha q_\alpha+\epsilon_\beta q_\beta-k}
\sigma^{\dagger}_{\epsilon_\alpha q_\alpha+\epsilon_\beta q_\beta-k}}
=\frac{4}{N}(N-n_e-2)(n_e-1) < 4n_e.
\end{equation}
Substituting the above inequality and Eq.~\eqref{skFq_4} into Eq.~\eqref{sigmaF}, we obtain that
\begin{equation}
\norm{\sigma^{\dagger}_{q_\alpha, q_\beta;k}\ket{F_{\vec{\xi}}}_{\alpha,\beta}}
<2\sqrt{n_e}\sum_{\epsilon_\alpha, \epsilon_\beta=\pm 1} \abs{h_k(\epsilon_\alpha q_\alpha, \epsilon_\beta q_\beta)}.
\end{equation}

Assembling the above results we obtain an inequality
\begin{equation}\label{bound}
\norm{\sigma_k \ket{F_{\vec{\xi}}}} < \frac{1}{N}\sum_{\alpha=1}^{n_e}  \abs{h_{k}(-q_{\xi_{\alpha}}, q_{\xi_\alpha})}
+\frac{2\sqrt{n_e}}{N}  \sum_{\substack{1\leq\alpha<\beta\leq n_e \\ \epsilon_{1},\epsilon_2=\pm 1}}
\abs{h_k(\epsilon_1 q_{\xi_\alpha}, \epsilon_2 q_{\xi_\beta})}.
\end{equation}
\end{widetext}

\subsection{Consistency of our perturbation theory approach}
The assumption $\xi_{n_e}\ll N$ implies that for $\abs{k}\sim N^{-1}$, the magnitude of $\sigma_k\ket{F_{\vec{\xi}}}$
scales as $O(1)$. This is because $\ket{k}$ overlaps with states in
the form of $\ket{\psi_{\xi_j}}$ that constitute $\ket{F_{\vec{\xi}}}$.

On the other hand, for $k$ not in the vicinity of $k_\ex=0$, i.e., $\abs{k}>p_0$ for any
constant  $p_0$ in the Brillouin zone, 
there will be a sufficiently large $N_0$ so that for $N>N_0$
we have $\norm{\sigma_k\ket{F_{\vec{\xi}}}}\sim N^{-2}$. To verify this,
in the previous section we have introduced
\begin{equation}
h_{k}(q_1,q_2)=\cot(\frac{k-q_{1}}{2}d)-\cot(\frac{k-q_2}{2}d).
\end{equation}
For $\abs{k}>p_0$ and $q_1, q_2\sim N^{-1}$ hence $q_1, q_2\ll \abs{k}$ for a sufficiently large $N$. 
Then the above expression is approximated by
\begin{equation}
\begin{aligned}
h_k(q_1, q_2)&\approx \frac{1}{\sin^2(kd/2)}\frac{q_1-q_2}{2}d \\
&\sim N^{-1}.
\end{aligned}
\end{equation}
The inequality derived above then leads to $\norm{\sigma_k\ket{F_{\vec{\xi}}}}\sim N^{-2}$, or
equivalently, $\braket{ F_{\vec{\xi}}| \sigma_k^\dagger \sigma_k | F_{\vec{\xi}}}\sim N^{-4}$.

In summary, we have
\begin{equation}
\braket{ F_{\vec{\xi}}| \sigma_k^\dagger \sigma_k | F_{\vec{\xi}}}\sim
\begin{cases} 
O(1), & \text{if}\;\; \abs{k}\sim N^{-1} \\
N^{-4}, & \text{if}\;\; \abs{k}> O(N^{-1})
\end{cases}
\end{equation}
Recall that we can rewrite $\Delta H$ as
\begin{equation}
\Delta H=N\int_{-\pi/d}^{\pi/d}\frac{dk}{2\pi}\;\delta\omega(k)\,\sigma_k^{\dagger}\sigma_k
\end{equation}
where $\delta\omega(k)=\omega_\eff(k)-\omega_1(k)$ can be given by a Taylor series approximation 
\begin{equation}
\delta\omega(k)=a_3 (k-k_\ex)^3+O(k^4),\;\text{for}\; k\approx k_\ex.
\end{equation}
Alternatively, we may use the Lagrange form of the remainder of 
the 2nd order Taylor polynomial, and write
\begin{equation}
\delta\omega(k)=\frac{\delta\omega^{(3)}(\xi_k)}{6}(k-k_\ex)^{3}
\end{equation}
where $\delta\omega^{(3)}$ is given by the 3rd order derivative of $\delta\omega$
and $\xi_k$ is a value in the interval between $k_{\ex}$ and $k$.

To show that $\Delta H$ is a perturbation to $\ket{F_{\vec{\xi}}}$ with $\xi_{n_e}\ll N$, we rewrite
Eq.~(8) of the main text here:
\begin{equation}
\braket{F_{\vec{\xi}}|\Delta H |F_{\vec{\xi}} } =
N\int_{-\pi/d}^{\pi/d}\frac{dk}{2\pi}\;\delta\omega(k)\,
\braket{F_{\vec{\xi}}|\sigma_k^{\dagger}\sigma_k |F_{\vec{\xi}} }.
\end{equation}
We notice that in a narrow interval of $k\sim N^{-1}$ (the measure
of this interval is $O(N^{-1})$)
we have $\delta\omega(k)\sim N^{-3}$ and $\braket{F_{\vec{q}}|\sigma_k^{\dagger}\sigma_k |F_{\vec{q}} }\sim O(N^0)$.
While for the integral of $k$ not in the vicinity of $k_\ex=0$, for which the measure is $O(1)$,
we have $\delta\omega(k)\sim O(1)$
but $\braket{F_{\vec{\xi}}|\sigma_k^{\dagger}\sigma_k |F_{\vec{\xi}} }\sim O(N^{-4})$. 
Therefore, the scaling of the integral must be
\begin{equation}
\braket{F_{\vec{\xi}}|\Delta H |F_{\vec{\xi}} }\sim N^{-3}.
\end{equation}
This shows that $\braket{F_{\vec{\xi}}|\Delta H |F_{\vec{\xi}} }$ will
be much smaller than the energy separation between different $\ket{F_{\vec{\xi}}}$, which scale as $N^{-2}$, and confirms the consistency of our perturbation approach.

\section{Derivation of Eq.~(9) of the main text}
\label{derive10}
The quantized electric field given in Eq.~(9) of the
main text can be formally derived from a Green tensor approach to the quantization of the
electromagnetic field with the presence of dispersive and dissipative 
materials~\cite{Gruner1996}. In this approach, a three-dimensional bosonic field 
$\mathbf{f}(\mathbf{r},\omega)$ is introduced. The field satisfies the standard
bosonic commutation relation 
\begin{equation}
    [f_i(\mathbf{r},\omega), f^\dagger_j(\mathbf{r'}, \omega']=\delta_{i,j}
    \delta(\mathbf{r}-\mathbf{r'})\delta(\omega-\omega')
\end{equation}
and $[f_i, f_j]=0$. The electric field is quantized as
\begin{equation}
    \begin{aligned}
    \mathbf{E}(\mathbf{r})= &i\mu_0\sqrt{\frac{\hbar\epsilon_0}{\pi}}\int_{0}^{\infty}d\tilde{\omega}\int d^3\mathbf{r'}\,\tilde{\omega}^2\sqrt{\Im \epsilon(\mathbf{r'},\tilde{\omega})} \\
    & \times \mathbf{G}(\mathbf{r},\mathbf{r'},\tilde{\omega})\cdot\mathbf{f}(\mathbf{r'},\tilde{\omega})+h.c.,
    \end{aligned}
\end{equation}
where $\Im \epsilon(\mathbf{r'},\tilde{\omega})$ is the imaginary part of the
relative permittivity. The free Hamiltonian of the electromagnetic field is given by
\begin{equation}
    H_0=\int_{0}^{\infty}d\omega\int d^3\mathbf{r}\;\hbar\omega 
    \mathbf{f}^{\dagger}\mathbf{f}(\mathbf{r},\omega).
\end{equation}
The interaction Hamiltonian between the emitters and the electromagnetic field is
\begin{equation}
    H_{int}=-\sum_{i=1}^{N} \sigma_{x,i}\mathbf{d}_i\cdot\mathbf{E}(\mathbf{r_i}),
\end{equation}
where $\mathbf{r_i}$ is the position of the i-th emitter, and we have assumed
a real-value transition dipole for the sake of convenience. 
Then the Heisenberg equation for the field operator $\mathbf{f}$ can be formally solved as
\begin{equation}
\begin{aligned}
    \mathbf{f}(\mathbf{r},\omega)[t]= &\mathbf{f}(\mathbf{r},\omega)[t_0]e^{-i\omega (t-t_0)}
    +\mu_0\sqrt{\frac{\epsilon_0}{\hbar\pi}\Im\epsilon(\mathbf{r},\omega)} \omega^2 \\
    \times \sum_{m=1}^{N} \mathbf{d_m} & \cdot\mathbf{G}^{*}(\mathbf{r},\mathbf{r_m},\omega)
    \int_{t_0}^{t}[\sigma_{m}(\tau)+\sigma_{m}^{\dagger}(\tau)]e^{-i\omega(t-\tau)}
\end{aligned}
\end{equation}
Here we ignore the contribution from the initial value, and use the Markov approximation
by substituting $\sigma_i(\tau)$ with $\sigma_i(t)e^{i\omega_0(t-\tau)}$, where $\omega_0$ is
the resonant frequency of the atomic transition. This leads to
\begin{equation}
\begin{aligned}
    \mathbf{f}(\mathbf{r},\omega)[t]=&
    \mu_0\sqrt{\frac{\epsilon_0}{\hbar\pi}\Im\epsilon(\mathbf{r},\omega)} 
    \omega^2 \sum_{m=1}^{N}\mathbf{d_m}\cdot\mathbf{G}^{*}(\mathbf{r},\mathbf{r_m},\omega)
    \\
   \times  & [
    \sigma_m(t)\frac{-i}{\omega-\omega_0-i0^{+}} +
    \sigma^{\dagger}_m(t)\frac{-i}{\omega+\omega_0-i0^{+}} ].
\end{aligned}
\end{equation}
A conjugate result can be obtained for $\mathbf{f}^{\dagger}[t]$.
Using the relation
\begin{equation}
\begin{aligned}
    & \frac{\omega^2}{c^2}\int d^3\mathbf{r} \Im\epsilon(\mathbf{r},\omega)
    G_{ik}(\mathbf{r_1},\mathbf{r},\omega)G^{*}_{jk}(\mathbf{r},\mathbf{r_2},\omega) \\
    = &  \Im G_{ij}(\mathbf{r_1},\mathbf{r_2},\omega),
\end{aligned}
\end{equation}
we obtain that
\begin{equation}
\begin{aligned}
\mathbf{E}(\mathbf{r})=& \mu_0\frac{1}{\pi}\sum_{m=1}^{N} \int_{0}^{\infty}d\omega (\mathbf{d}_m)_k \Im G_{ik}(\mathbf{r}, \mathbf{r_m},\omega )\sigma_m(t) \\
&\times  \omega^2
[\frac{1}{\omega-\omega_0-i0^{+}}+\frac{1}{\omega+\omega_0+i0^{+}}]+h.c.
\end{aligned}
\end{equation}
Therein, terms in the square bracket are equal to
\begin{equation}
    \mathcal{P}\frac{2\omega}{\omega^2-\omega_0^2}
    +i\pi\delta(\omega-\omega_0)-i\pi\delta(\omega+\omega_0),
\end{equation}
where $\mathcal{P}$ denotes the principal value. 
We recall the Kramers-Kronig relation for a function $\chi(\omega)$, which is 
analytic in the closed upper half plane of complex $\omega$ and vanishes like $1/\abs{\omega}$ or
faster as $\abs{\omega}\rightarrow \infty$ and $\chi(\omega)=\chi^*(-\omega)$
\begin{equation}
    \Re\chi(\omega)=\frac{2}{\pi}\int_{0}^{\infty}d\omega'\mathcal{P}\frac{\omega' \Im \chi(\omega')}{\omega'^2-\omega^2}.
\end{equation}
Substituting $\chi(\omega)$ with $\omega^2\mathbf{G}(\omega)$, we obtain that
\begin{equation}
     \frac{1}{\pi}\int_{0}^{\infty}d\omega\, 
     \Im G_{ik}(\mathbf{r}, \mathbf{r_m},\omega )
     \mathcal{P}\frac{2\omega^3}{\omega^2-\omega_0^2}=\omega_0^2\Re G_{ik}(\mathbf{r}. \mathbf{r_m},\omega_0 )
\end{equation}
Then Eq.~(9) of the main text can be obtained immediately.

\section{Atom array in the 3D free space}
\label{scheme1}

The RDDI Hamiltonian of an atom array in 3D free space has a closed form expression. It is used in the detection scheme-1 and in
detection scheme-2 after supplementing its dissipative part with a scaling factor $\beta$. 

\begin{figure}[bt]
\centering
\includegraphics[width=0.9\textwidth]{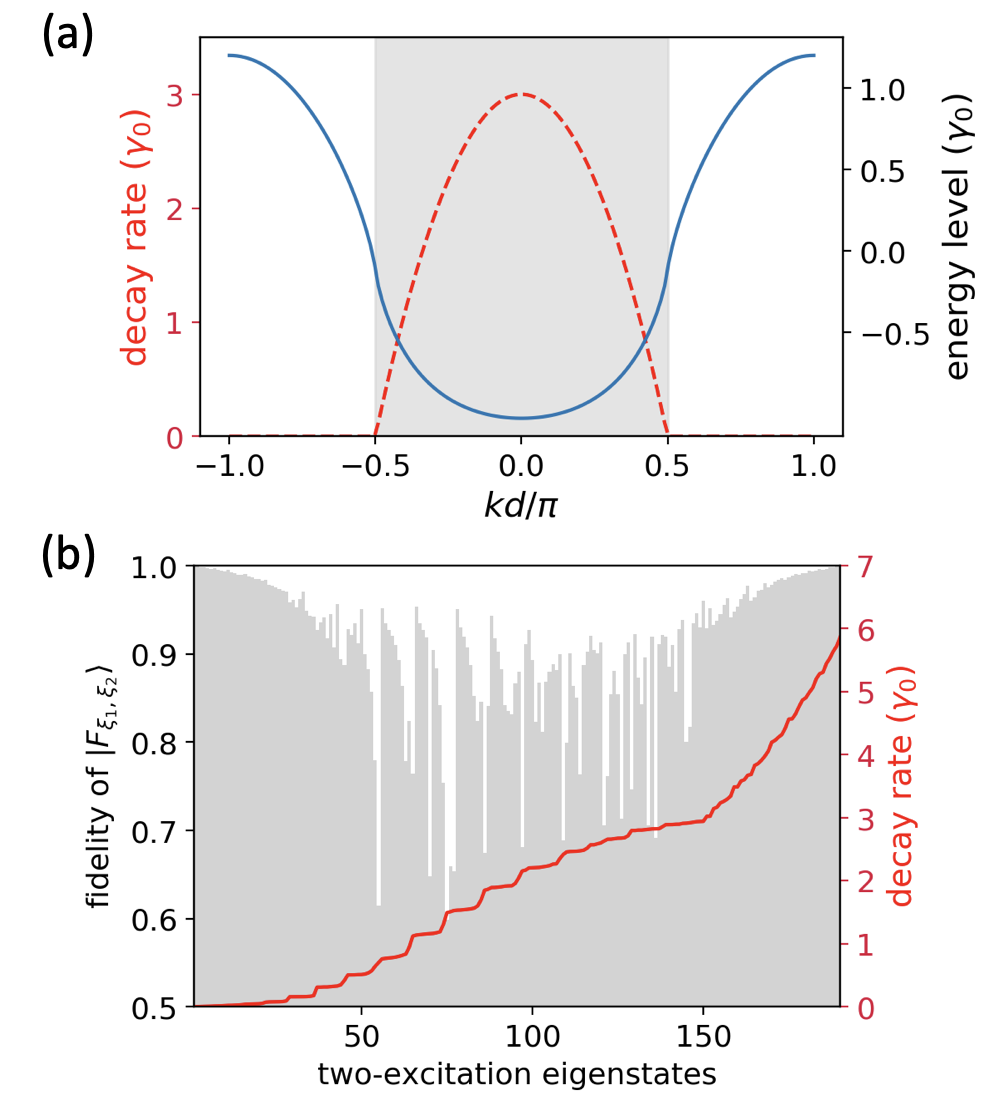}
\caption{ (a) $\Re\omega_\eff(k)$ (blue curve) and $\gamma(k)=-2\Im\omega_\eff(k)$ (red curve) 
for 1D atom array in 3D free space, with
the atomic transition dipoles oriented parallel to the array direction. $\gamma(k)=0$ outside the shaded light cone region. (b) State fidelities (grey bars) with the free-fermion ansatz of all the two-excitation
eigenstates of an array with $N=20$ atoms. The eigenstates are found by numerical diagonalization and  sorted according to  increasing decay rates (red curve).
}\label{spfig1}
\end{figure}

The dyadic Green's tensor is given by
\begin{equation}
\begin{aligned}
\mathbf{G}(\mathrm{r}, \omega_0)= & \frac{e^{ik_0 r}}{4\pi k_0^2 r^3}\bigg[ (k_0^2 r^2+ik_0 r-1)\mathbb{I}+\\
&\; (-k_0^2 r^2-3ik_0 r+3)\frac{\mathbf{rr}}{r^2} \bigg],
\end{aligned}
\end{equation}
where $\mathbb{I}$ is the $3\times 3$ identity tensor, $\mathbf{rr}$ is the dyadic vector product, and $r$ is the magnitude of $\mathbf{r}$. As in the main text we
suppose the atomic transition dipoles are aligned parallel to the array direction. Then the
dispersion relation is given by
\begin{equation}
\omega_\eff(k)=-i\frac{3\gamma_0}{2}\sum_{\substack{\epsilon=\pm 1 \\ \xi=2,3}}
(\frac{i}{k_0 d})^{\xi} \mathrm{Li}_{\xi}[e^{i(k_0+\epsilon k)d}]
\end{equation}
where $\mathrm{Li}_{\xi}(z)=\sum_{n=1}^{\infty} z^n n^{-\xi}$ is the polylogarithm of order $\xi$. 
The second order derivative of $\omega_{\eff}(k)$ is
\begin{equation}
\frac{d^2\omega_\eff(k)}{dk^2}=i\frac{3d^2\gamma_0}{2}\sum_{\substack{\epsilon=\pm 1 \\ \xi=0,1}}
(\frac{i}{k_0 d})^{\xi+2} \mathrm{Li}_{\xi}[e^{i(k_0+\epsilon k)d}]
\end{equation}
where $\mathrm{Li}_0(z)=\frac{z}{1-z}$ and $\mathrm{Li}_1(z)=-\ln (1-z)$. 
We plot $\Re\omega_\eff(k)$ (blue curve) and the decay rate $-2\Im\omega_\eff(k)$ (red curve) in 
Fig.~\ref{spfig1}(a), for atom arrays with $k_0 d=\pi/2$. It can be seen that $k=0$ and $k=\pi/d$ are
extremum points. 

We numerically obtain the two-excitation eigenstates of the effective RDDI Hamiltonian for an
array of $N=20$ atoms, and
show their fidelities with the free-fermion states (grey bars) in Fig.~\ref{spfig1}(b), where the states are sorted by increasing decay rates (red curve). 
Formally, the shown fidelity is defined as
\begin{equation}
\max_{\xi_1, \xi_2}\abs{\braket{F_{\xi_1, \xi_2}|\psi_j}}^2
\end{equation}
for all 190 two-excitation eigenstates $1\leq j\leq 190$. It confirms that there are two families of states around the most subradiant and the most superradiant states that
have high fidelities with the free-fermion ansatz.

\begin{figure}[b]
\centering
\includegraphics[width=0.9\textwidth]{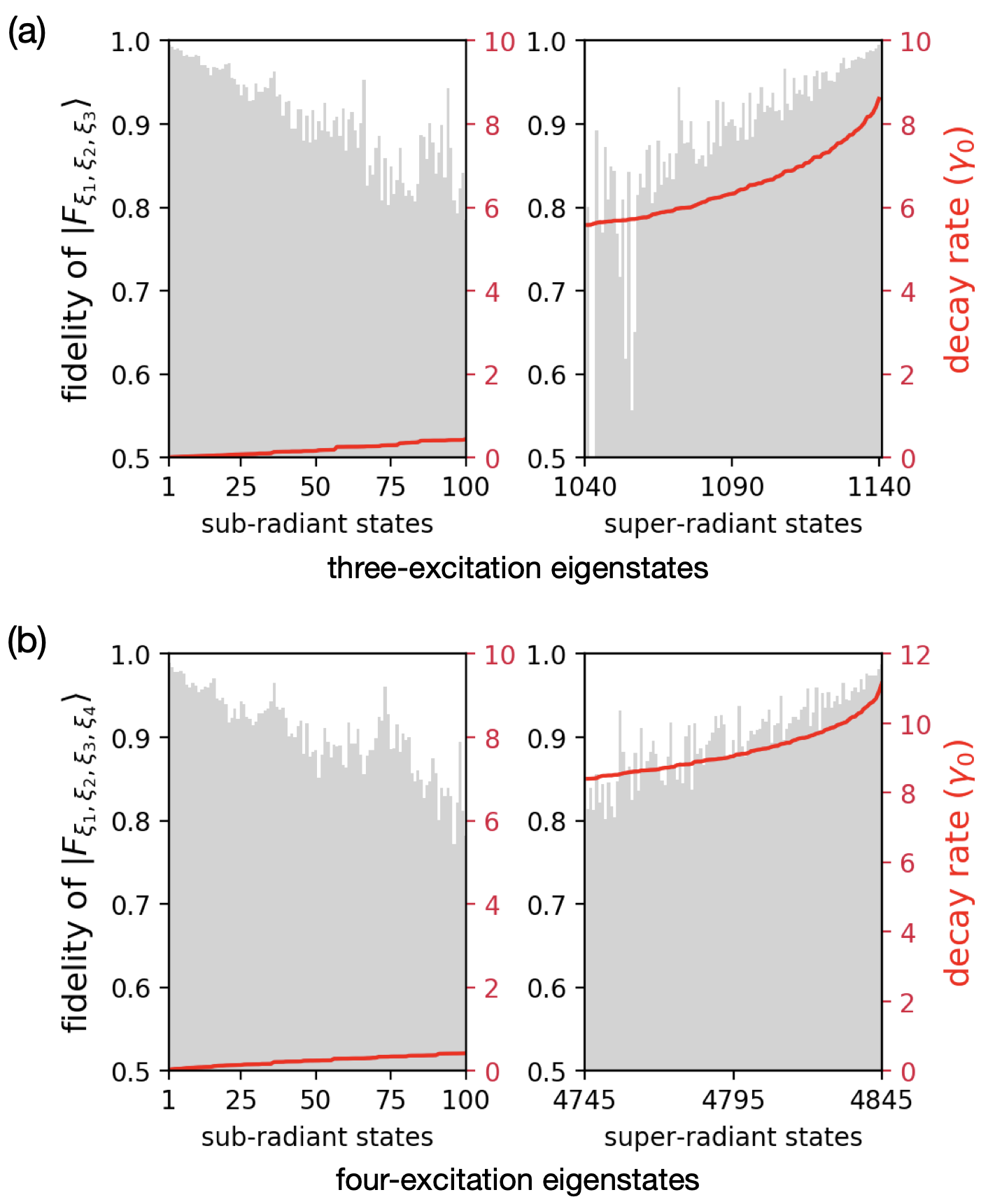}
\caption{ 
State fidelities with the free-fermion ansatz of the
(a) three-excitation and (b) four-excitation eigenstates of the system same with 
Fig.~\ref{spfig1}(b). The eigenstates are sorted according to increasing decay rates (red curves).
In each sub-figure, the left panel is plotted for
the 100 most sub-radiant states, and the right panel is plotted for the 100 most super-radiant states.}
\label{spfig1e34}
\end{figure}

Next, we extend the same calculations to the three-excitation and four-excitation eigenstates.
The array of $N=20$ atoms has 1140 and 4845 three- and four-excitation eigenstates, respectively.
We pick up the 100 most sub-radiant and 100 most super-radiant states, corresponding to the states
near $k_{\ex,0}$ and $k_{\ex,\pi}$, respectively, and
plot the fidelities of the free-fermion ansatz in Fig.~\eqref{spfig1e34}. 
On one hand, it confirms that the free-fermion states are good approximations.
On the other hand, it shows that the highest fidelity (obtained at the first and the last state, sorted by increasing decay rates) decreases with the 
increasing number of excitations.

\section{Simulations of Scheme-2 with $N=30$}
\label{scheme2}

It is argued that the ratio $r_\beta=\Delta\omega/(\beta\gamma_{\ex})$ with $\Delta\omega=\Re a_2/(Nd)^2$
determines the visibility of the free-fermion feature. However,
in the main text,  results are presented only for $N=20$. Calculations with identical parameters but $N=30$ show that a larger $N$ leads to a smaller
energy gap between the eigenstates and hence blurs the visibility of the free-fermion states.  
The simulation is restricted to the Hilbert space of $n_e\leq 3$, and obtained by averaging over
150 samplings of Monte Carlo quantum trajectories.

\begin{figure}[hb]
\centering
\includegraphics[width=1\textwidth]{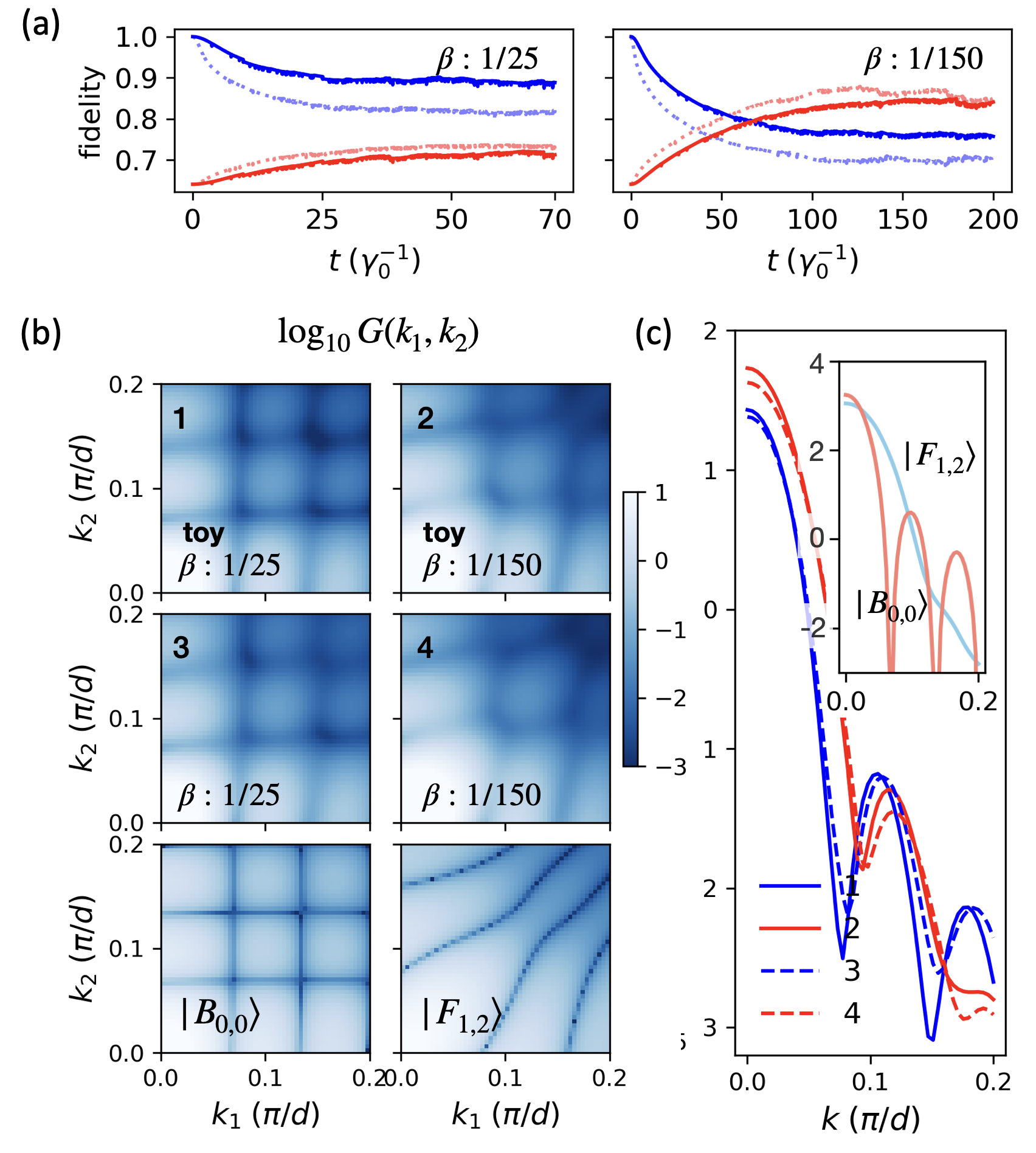}
\caption{Detection scheme-2 with $N=30$. (a) Fidelities between the two-excitation components
of the steady states (normalized) and the bosonic state $\ket{B_{0,0}}$ (blue) and $\ket{F_{1,2}}$ (red), with solid lines
and dotted lines defined as in Fig.~2 of the main text.
(b) $\log_{10}G(k_1, k_2)$ evaluated from the steady states, with the three rows defined as in Fig.~2
of the main text. The color map belongs to the 1st and 2nd rows. The color map of the 3rd row is not shown.
(c) $\log_{10}G(k, k)$, i.e., the diagonal lines of the plot in (b). 
}\label{spfig2}
\end{figure}

Comparing Fig.~\ref{spfig2}(a) with Fig.~(2a) of the main text, we see that now for $\beta=1/25$ the steady states are
apparently closer to the free-boson state $\ket{B_{0,0}}$; and for $\beta=1/150$, the steady states are closer to but not perfectly described by
 $\ket{F_{1,2}}$.  As in Fig.~(2a) of the main text, the red solid and dotted curves match
better than the blue curves. This is because $\ket{B_{0,0}}$ has overlap with states $\ket{F_{\xi_1,\xi_2}}$ that are
not close to $k_\ex=0$, for which our ``toy model'' becomes less precise. 

In Fig.~\ref{spfig2}(b) we plot $\log_{10}G(k_1, k_2)$ for $0\leq k_1, k_2\leq 0.2\pi/d$. The arrangement of the
subplots is same as in the main text. It can be seen that patterns of suppressed coincidence are different for  $\ket{B_{0,0}}$
and $\ket{F_{1,2}}$: the former is featured by four upright crossings, which are absent in the latter.
The two states of $\beta=1/25$ clearly display the crossings, while for $\beta=1/150$ the crossings are deformed.
Comparing them with the plot of $\ket{F_{1,2}}$ we can see that the character of the free-fermion state
is not clear, demonstrating that a larger $N$ reduces the visibility. 

In Fig.~\ref{spfig2}(c) we plot the diagonal lines of the subplots of Fig.~\ref{spfig2}(b). In the insert we see
that the blue solid and dashed lines for $\ket{B_{0,0}}$ feature two regions of suppressed coincidences
($\beta=1/25$). The curves for $\beta=1/150$ (red lines)  unambiguously show the first reduced coincidence, at about $k\approx 0.1\pi/d$, a little bit towards larger $k$ 
compared with those of $\ket{B_{0,0}}$ and the values for $\beta=1/25$. However, the second region with suppressed coincidences  is flattened for $\ket{F_{1,2}}$ (the red curves).

It is important to emphasize that the eigenstates of the system become better and better approximated by the  fermionic states for large $N$. What we have shown here is only that the reduced level spacing for larger $N$ makes it more difficult to prepare and detect the fermionic  states by optical means.

\bibliography{supsub}